%% file: crossing9.tex
\documentclass[12pt]{iopart}
\usepackage{amssymb}
\usepackage{epsfig}
\usepackage{subfigure}
\usepackage{graphicx,color}
\graphicspath{ {./figs/} }
\usepackage{textcomp}
\usepackage{cite}
\usepackage{float}

\expandafter\let\csname equation*\endcsname\relax
\expandafter\let\csname endequation*\endcsname\relax
\usepackage{amsmath}

\begin{document}
\title{Intermediate-Level Crossings of a First-Passage Path}

\author{Uttam Bhat} 
\address{Department of Physics, Boston University, Boston, MA 02215, USA}
\address{Santa Fe Institute, 1399 Hyde Park Road, Santa Fe, NM 87501, USA}
\author{S. Redner}
\address{Santa Fe Institute, 1399 Hyde Park Road, Santa Fe, NM 87501, USA}
\address{Center for Polymer Studies and Department of Physics, Boston University, Boston, MA 02215, USA}

\begin{abstract}

  We investigate some simple and surprising properties of a one-dimensional
  Brownian trajectory with diffusion coefficient $D$ that starts at the
  origin and reaches $X$ either: (i) at time $T$ or (ii) for the first time
  at time $T$.  We determine the most likely location of the first-passage
  trajectory from $(0,0)$ to $(X,T)$ and its distribution at any intermediate
  time $t<T$.  A first-passage path typically starts out by being repelled
  from its final location when $X^2/DT\ll 1$.  We also determine the
  distribution of times when the trajectory first crosses and last crosses an
  arbitrary intermediate position $x<X$.  The distribution of first-crossing
  times may be unimodal or bimodal, depending on whether $X^2/DT\ll 1$ or
  $X^2/DT\gg 1$.  The form of the first-crossing probability in the bimodal
  regime is qualitatively similar to, but more singular than, the well-known
  arcsine law.

\end{abstract}
\pacs{05.40.Jc, 02.50.-r, 05.40.-a}
\maketitle

\section{Introduction}

While Brownian motion has been extensively studied and much is well
understood about this process (see, e.g., \cite{L39,L65,F68,MP10}), a number
of simple questions continue to be investigated when the motion is subject to
a constraint.  Some examples of this genre include determining the time that
Brownian path spends in a range $dx$ about a point $x$~\cite{DK57,R63,K63},
the area swept out by a Brownian motion when it first returns to the
origin~\cite{C75,C76,MC04,MC05}, the time when a one-dimensional Brownian
motion, which starts at $x>0$, attains its maximum before first crossing the
origin~\cite{RM07,MRKY08}, as well as related extremal properties of such
motions~\cite{PCMS11}.  There has also been considerable study on the topic
of level-crossings, namely, the crossings of a given point on the line, for
both Brownian motion~\cite{A02,B00} and for general stochastic
processes~\cite{K06,PY86,ASG93}, as well as investigations of the crossings
of a time-dependent level (see, e.g.,~\cite{L65,A02,L86,B67,G89,NFK99,D00}).

\begin{figure}[ht]
\centerline{\resizebox{0.4\textwidth}{!}{\input{figs/model.pspdftex}}}
\caption{Illustration of the quantities under study.  Shown is a schematic
  first-passage path from $(0,0)$ to $(X, T)$.  At some intermediate time
  (dashed line), the particle is at position $x$.  At some intermediate level
  (dotted line), the first- and last-crossing times of this level are $t_f$
  and $t_\ell$. }
\label{model}
\end{figure}
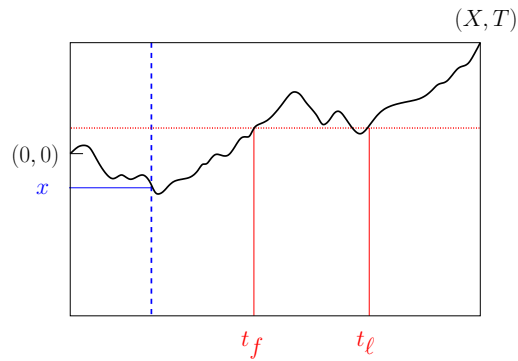

Given these extensive studies, it is surprising that some basic issues about
level crossings seem to have not yet been fully addressed.  In this work, we
investigate the intermediate-level crossings in both space and time of a
Brownian motion that \emph{first} reaches a final destination $X>0$ at time
$T$.  This type of question underlies the task of reconstructing a Brownian
path from measurements at discrete times, an issue that is of basic
importance for understanding diffusion in porous media (see, e.g.,
\cite{MSSL92,SSMH94,ZJS11}).  The condition that the particle must reach $X$
for the first time at time $T$ constrains the entire trajectory before time
$T$ in a non-trivial way.  We will study the consequences of this constraint
on the following basic properties of a Brownian particle with diffusivity $D$
at intermediate stages on its way to its first passage at $(X,T)$, as
illustrated in Fig.~\ref{model}:
\begin{enumerate}

\item At an arbitrary intermediate time, what is the expected position $x$ of
  the particle?

\item When does the particle \emph{first} cross an intermediate level $x<X$?

\item When does the particle \emph{last} cross this intermediate level?

\end{enumerate}

Our approach to answer these questions is based on decomposing the
first-passage probability to $(X,T)$ as the convolution of the propagator to
the intermediate point $(x,t)$ and the propagator from this intermediate to
the final point.  In this framework, the intermediate crossing properties are
simple to formulate; however, the results are somewhat surprising and depend
in an important way on the global nature of the full first-passage path.  In
the ``ballistic limit'', where $X^2\gg DT$, the space-time trajectory of the
first-passage path approaches  a straight line.  In this case,
intermediate-crossing properties are close to those that arise by treating
the first-passage trajectory as purely ballistic.  In the opposite
``diffusive limit'', where $X^2\ll DT$, the first-passage path is typically
repelled from its final location $X$ so as to avoid hitting $X$ before time
$T$.

Correspondingly, the probability distribution of times for the particle to
first cross an intermediate level $0<x<X$ undergoes a transition from
unimodality, for $X^2\gg DT$, to bimodality, for $X^2\ll DT$.  In the latter
case, the first-passage path to $(X,T)$ is most likely to cross an
intermediate level near the beginning or near the end of its trajectory.  In
this bimodal regime, the distribution of the first-crossing times is
reminiscent of, but more singular than, the famous arcsine
law~\cite{F68,MP10} for the distribution of times that a one-dimension
Brownian particle spends on one side of the origin, as well as the
distribution of the last passage time at the origin.

As a preliminary, we first investigate intermediate crossing phenomena for a
purely Brownian path between $(0,0)$ and $(X,T)$ in section~\ref{sec:prelim}.
In spite of the simplicity of this problem, the distributions of the first-
and last-crossing times of an intermediate $0<x<X$ level are also quite rich
and have a qualitatively similar unimodal to bimodal transition as in the
case where the trajectory is a first-passage path.  Moreover, the boundary in
phase space that demarcates this transition is, surprisingly, not single
valued.  In section~\ref{sec:FPP}, we investigate intermediate crossing
properties when the trajectory between $(0,0)$ and $(X,T)$ is a first-passage
path.  We briefly discuss the situation where the intermediate level is
negative in section~\ref{sec:neg}.  Finally, we offer some perspectives in
section~\ref{sec:summ}.
 
\section{Preliminaries: Intermediate Crossings for Unconstrained Brownian Motion}
\label{sec:prelim}

Consider a Brownian particle in one dimension that starts at the origin.  We
will make extensive use of two important characteristics of this Brownian
motion: (i) the occupation probability and (ii) the first-passage
probability.  These quantities are, respectively:
\begin{equation}
\label{PF}
P(X,T) = \frac{1}{\sqrt{4\pi DT}}\,\, e^{-X^2/4DT}\,\qquad \qquad 
F(X,T) = \frac{\left\vert X\right\vert }{\sqrt{4\pi DT^3}}\,\,e^{-X^2/4DT}\,.
\end{equation}
Here $P(X,T)$ is the probability that a Brownian particle moves a distance
$X$ from its starting point over a time $T$, while $F(X,T)$ is the
probability that the particle \emph{first} reaches this point at time
$T$~\cite{R01}.  One of the intriguing aspect of a Brownian trajectory in one
dimension is that it ultimately reaches any point on the infinite line.
However, even though the trajectory is guaranteed to eventually reach any
point $X\ne 0$, the mean time to reach an arbitrary point is infinite.  This
dichotomy arises because the probability that the trajectory first reaches
position $X$ at time $T$ asymptotically decays as $T^{-3/2}$, from which the
mean time to reach $X$ is infinite.

For a Brownian particle that starts at $(0,0)$ and ends at $(X,T)$, we now
ask: where is the particle and what is its probability distribution of
positions at an earlier time $t<T$?  The probability that the particle
propagates from $(0,0)$ to $(X,T)$ may be written as the convolution of the
probabilities for the path from $(0,0)$ to $(x,t)$ and the path from $(x,t)$
to the final location $(X,T)$:
\begin{align}
\label{conv}
P(X,T) &=  \int P(x,t)\,\,
P(X\!-\!x,T-t)\,  dx \nonumber\\
&= \int \frac{e^{-x^2/4Dt}}{\sqrt{4\pi Dt}}\,\,\,
 \frac{ e^{-(X-x)^2/4D(T\!-\!t)}}{\sqrt{4\pi D(T\!-\!t)}}\,\,  dx\,.
\end{align}
Integrating over $x$ reproduces, after some simple algebra, $P(X,t)$ in
Eq.~\eqref{PF}.  For later convenience we introduce the dimensionless time
$\tau\!=\!t/T$ and the dimensionless length $\chi\!=\!x/X$.  By maximizing
the integrand with respect to $x$, it is straightforward to derive that the
integrand has its maximum when $x= \tau X$.  That is, the particle moves a
fraction $\tau$ of the final distance in a fraction $\tau$ of the total time,
a well-known property of Brownian motion~\cite{E83}.

Finally, we compute the probability $\mathcal{P}(x,t)$ that the particle is
at $x$ at time $t$, given that it is at $X$ at time $T$.  This quantity
equals the probability for the subset of Brownian paths from $(0,0)$ to
$(X,T)$ that reach $x$ at time $t$ divided by the probability for all
Brownian paths from $(0,0)$ to $(X,T)$:
\begin{align}
\label{CP}
\mathcal{P}(x,t)&= \frac{ P(x,t)\,\,
P(X\!-\!x,T\!-\!t)}{P(X,T)}\nonumber \\
&=\frac{ e^{-x^2/4Dt}}{\sqrt{4\pi Dt}}\,\,\,\,
 \frac{e^{-(X-x)^2/4D(T\!-\!t)}}{\sqrt{4\pi D(T\!-\!t)}}\, 
\times \sqrt{4\pi DT} \, e^{X^2/4DT} \nonumber \\
&= \frac{ e^{-(x-\tau X)^2/4DT \tau(1-\tau)}}{\sqrt{4\pi DT  \tau(1-\tau)}}\nonumber\\
&\equiv \sqrt{\frac{\alpha}{\pi}}\,\,\,
\frac{ e^{-\alpha (\chi-\tau)^2/\tau(1-\tau)}}{\sqrt{\tau(1-\tau)}}\,,
\end{align}
where we have rewritten the last line in terms of $\chi\! \in|! (-\infty,1)$,
$\tau\!\in\! (0,1)$, and the dimensionless parameter $\alpha= X^2/4DT$, which
demarcates whether the full trajectory is ballistic, for $\alpha\gg 1$, or
diffusive, for $\alpha\ll 1$.

From Eq.~\eqref{CP}, the probability distribution to reach an intermediate
point $x$ at dimensionless time $\tau$ is a Gaussian centered at $\chi=\tau$
whose width $\sqrt{\tau(1-\tau)}$ is maximal for $\tau=\frac{1}{2}$ and
vanishes for $\tau\to 0$ and $\tau\to 1$.  The physical message from these
results is that the average position of a Brownian particle that starts at
$(0,0)$ and ends at $(X,T)$ interpolates linearly between 0 and $X$ as the
time increases from 0 to $T$~\cite{E83}.

In a similar vein, we now determine when the Brownian particle \emph{first}
crosses and \emph{last} crosses an intermediate position $x<X$, given that
the particle is at $X$ (not necessarily for the first time) at time $T$.
More generally, we compute the probability distributions for these two
events.  Using the same reasoning that led to Eq.~\eqref{CP}, the probability
$\mathcal{F}(x,t)$ that the particle \emph{first crosses} $x$ at time $t$ is
given by:
\begin{subequations}
\label{CFL}
\begin{align}
\label{CF}
\mathcal{F}(x,t) &= \frac{F(x,t)\,P(X-x,T-t)}{P(X,T)}\,.
\end{align}
That is, the first-crossing probability equals the probability for a Brownian
particle, which starts at $(0,0)$, to \emph{first} reach $(x,t)$---hence the
factor $F(x,t)$---times the probability that the particle propagates from
$(x,t)$ to $(X,T)$, normalized by the probability for all Brownian paths that
propagate from $(0,0)$ to $(X,T)$.  

Similarly, the probability $\mathcal{L}(x,t)$ that the \emph{last crossing}
of $x$ occurs at time $t$ is:
\begin{align}
\label{CL}
\mathcal{L}(x,t) &= \frac{P(x,t)\,F(X-x,T-t)}{P(X,T)}\,.
\end{align}
\end{subequations}
The reasoning that underlies~\eqref{CL} is slightly more involved than that
for~\eqref{CF}.  For $(x,t)$ to be the last crossing, the remaining
trajectory must be a time-reversed first-passage path from $(X,T)$ to
$(x,t)$.  This constraint guarantees that the crossing at $x$ is the last one
in the full trajectory to $(X,T)$.

\begin{figure}[ht]
\centerline{\includegraphics[width=0.9\textwidth]{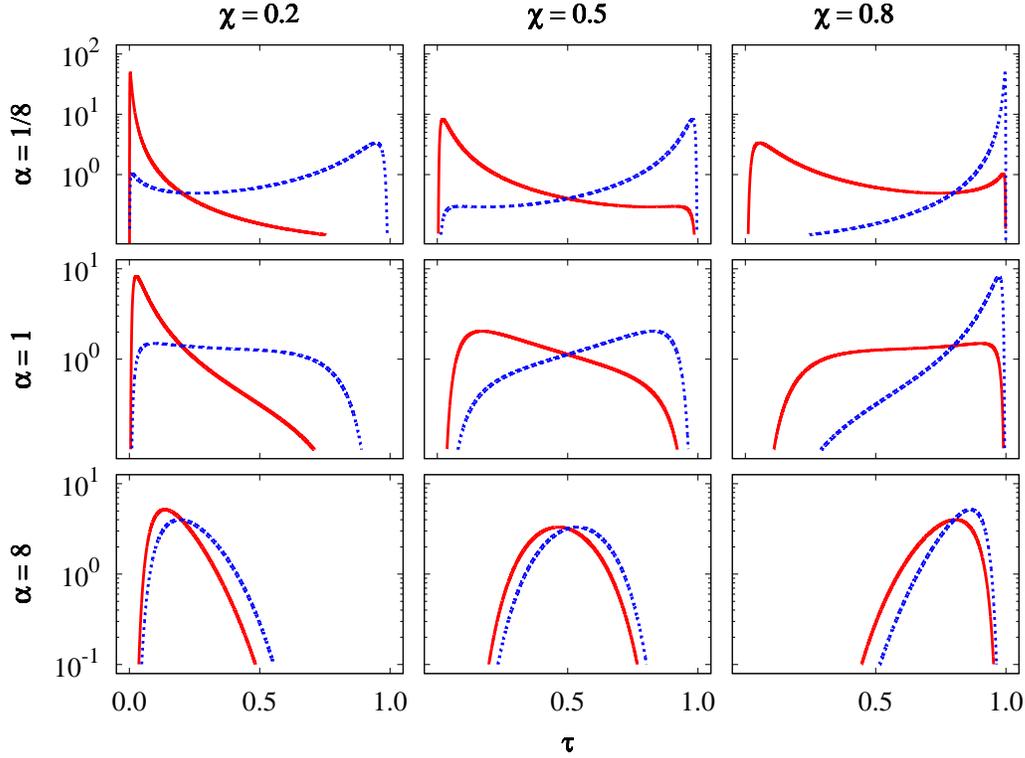}}
\caption{First-crossing probability $\mathcal{F}(\chi,\tau)$ (red, solid) and
  last-crossing probability $\mathcal{L}(\chi,\tau)$ (blue, dashed) for
  illustrative values of $\alpha=X^2/4DT$ and $\chi=x/X$ when the initial
  trajectory is an otherwise unrestricted Brownian path from $(0,0)$ to
  $(X,T)$.  For small $\alpha$, $\mathcal{F}(\chi,\tau)$ changes from
  unimodal to bimodal as $\chi$ increases.  For large $\alpha$ (almost
  ballistic trajectory), the first-crossing and last-crossing times are close
  to each other.}
\label{CF-CL-1}
\end{figure}

Substituting the explicit forms for the occupation and first-passage
probabilities from Eqs.~\eqref{PF} and performing some straightforward
algebra, the first-crossing and last-crossing probabilities are, after
expressing all variables in dimensionless form:
\begin{align}
\begin{split}
\label{CFLe}
\mathcal{F}(\chi,\tau) &=\sqrt{\frac{\alpha}{\pi}}  \frac{\left\vert \chi\right\vert }{\sqrt{\tau^3(1-\tau)}}
\,\, e^{-\alpha(\chi-\tau)^2/[\tau(1-\tau)]}\,,\\
\mathcal{L}(\chi,\tau) &= \sqrt{\frac{\alpha}{\pi}} \frac{\left\vert 1-\chi\right\vert }{\sqrt{\tau(1-\tau)^3}}
\,\, e^{-\alpha(\chi-\tau)^2/[\tau(1-\tau)]}\,.
\end{split}
\end{align}
It is apparent that
$\mathcal{F}(\chi,\tau)=\mathcal{L}(1\!-\!\chi,1\!-\!\tau)$, and vice versa,
so that we only need to study one of these quantities.  Figure~\ref{CF-CL-1}
shows the time dependence of the first-crossing probability
$\mathcal{F}(\chi,\tau)$ for three representative values of $\alpha$ and
$\chi$.  The first-crossing probability clearly becomes bimodal for small
$\alpha$ and $\chi\to 1$.  On the other hand, for $\alpha\to\infty$, so that
the trajectory becomes progressively more ballistic, the first- and
last-crossing probabilities must approach each other.

\begin{figure}[ht]
\subfigure[]{\includegraphics[width=0.48\textwidth]{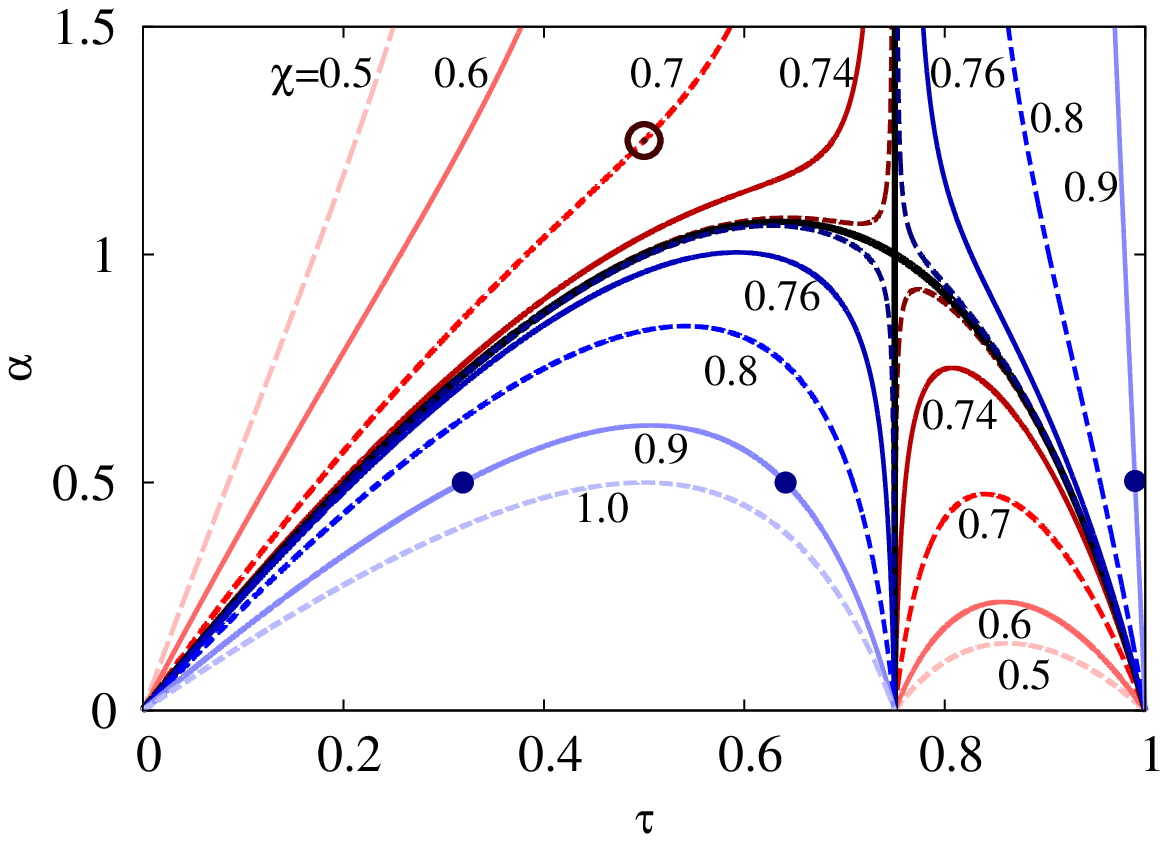}}
\subfigure[]{\includegraphics[width=0.48\textwidth]{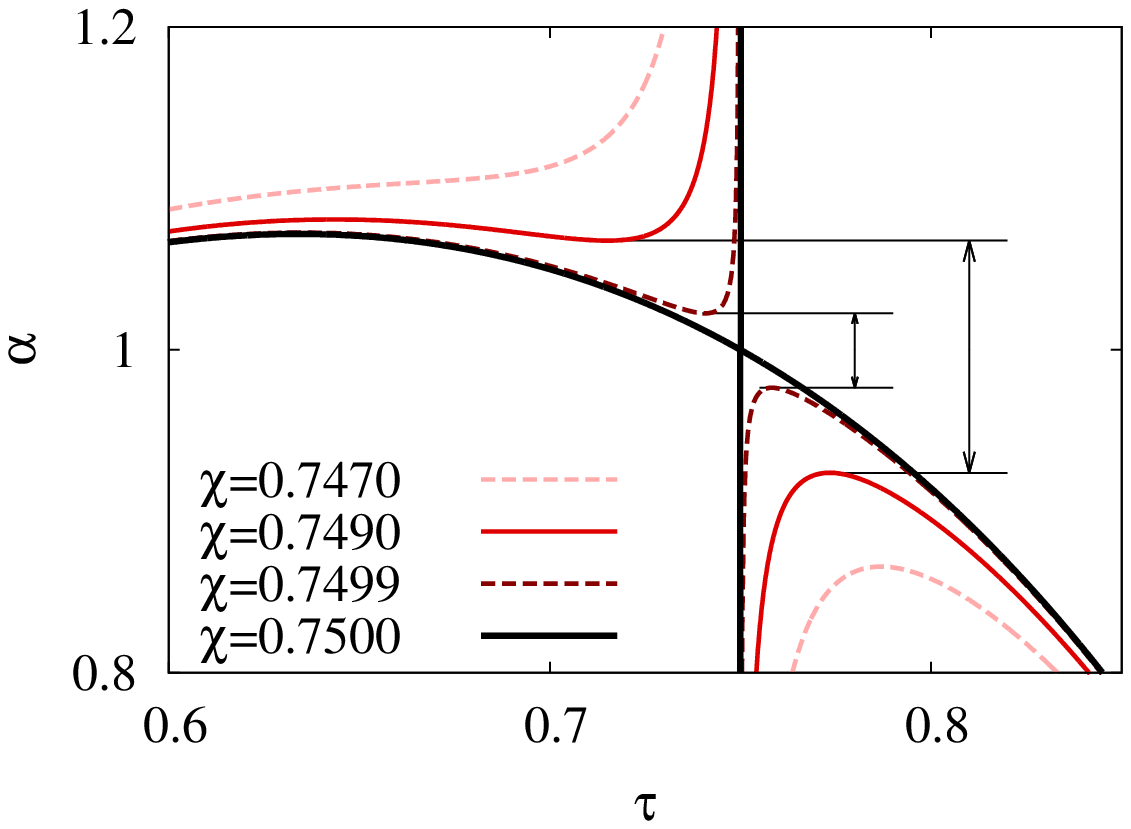}}
\caption{(a) Loci of the extrema of $\mathcal{F}(\chi,\tau)$ in the
  $\alpha$-$\tau$ plane for various values of $\chi$ when the initial
  trajectory is an otherwise unrestricted Brownian path to $(X,T)$.  For a
  given $\alpha$ and $\chi$, the extrema occur at the values of $\tau$ where
  the line $\alpha=\mathrm{const.}$ intersects the curve that corresponds to
  a given $\chi$. (b) Zoom near the critical point $(\frac{3}{4},1)$ }
\label{Atf}
\end{figure}

\begin{figure}[ht]
\subfigure[]{\includegraphics[width=0.4\textwidth]{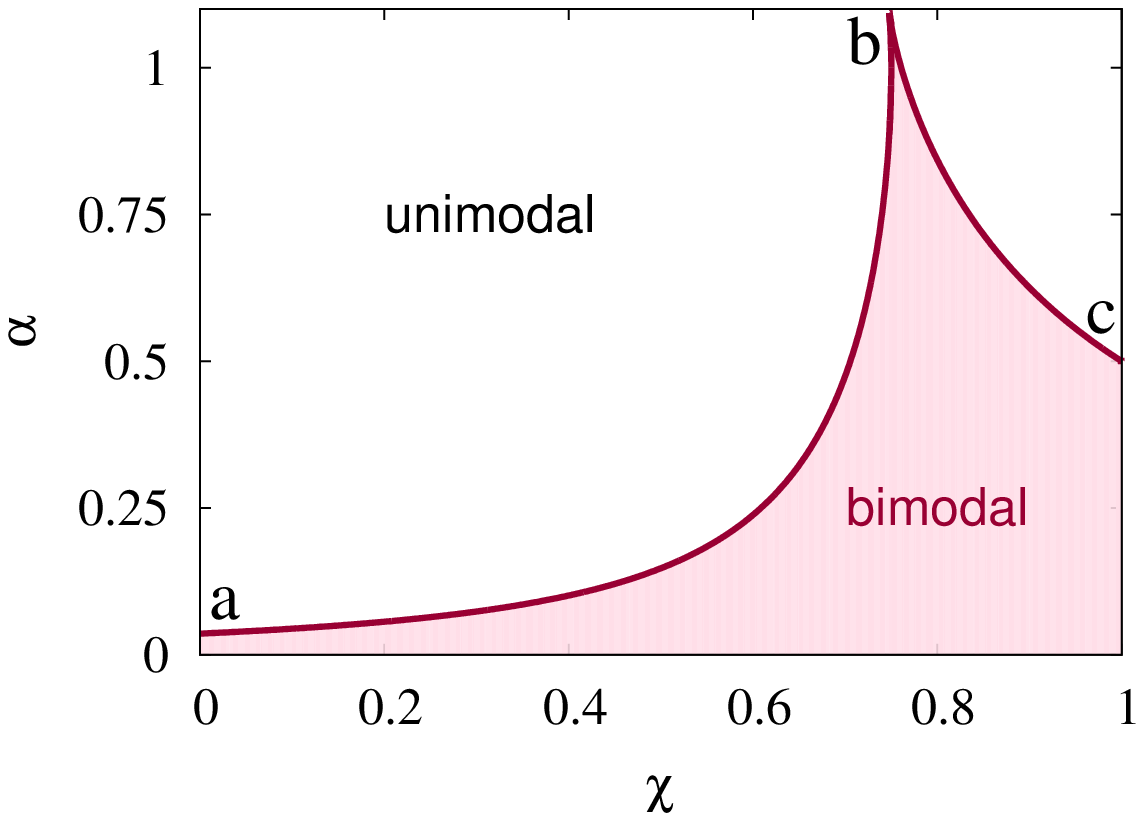}}
\subfigure[]{\includegraphics[width=0.4\textwidth]{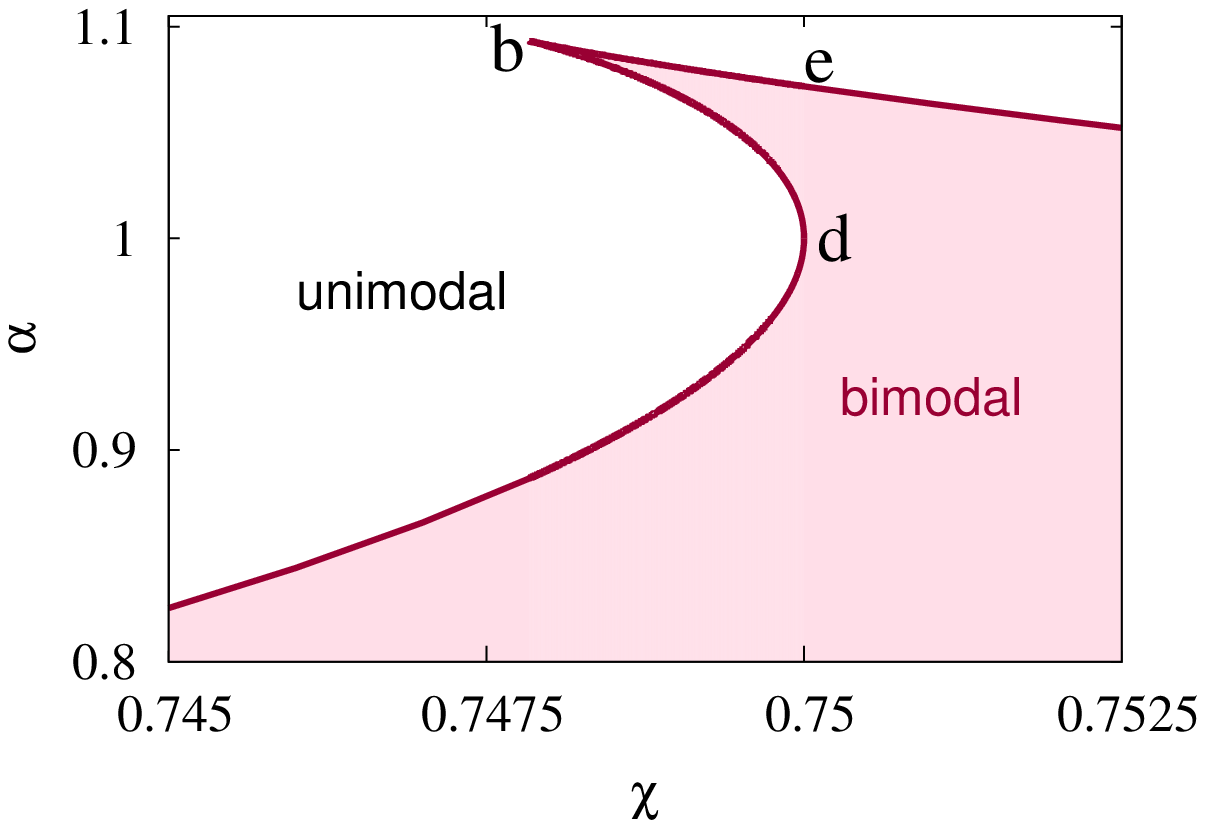}}
\caption{(a) Numerically obtained phase diagram from Eq.~\eqref{extr},
  showing the domains in the parameter space where the first-crossing
  probability is unimodal and where it is bimodal for an unrestricted
  Brownian path from $(0,0)$ to $(X,T)$.  The points marked $a,b,c$ are at
  $(0,\frac{7}{2}\!-\!2\sqrt{3})$, $(0.747835\cdots,1.09325\ldots)$, and
  $(1,\frac{1}{2})$, respectively.  (b) Detail near the cusp in the phase
  boundary; the points marked $d,e$ are at $(\frac{3}{4},1)$ and
  $(\frac{3}{4},4(2\!-\!\sqrt{3}))$, respectively.}
\label{Ac-1st-free}
\end{figure}

To determine where in the $\chi$-$\alpha$ plane that the first-crossing
probability changes from unimodal to bimodal, we first need to locate the
extrema of $\mathcal{F}(\chi,\tau)$ for each value of $\chi$.
Differentiating $\mathcal{F}(\chi,\tau)$ with respect to $\tau$ and setting
the result to zero gives the cubic equation
\begin{subequations}
\label{extr}
\begin{equation}
4\tau^3 +(2\alpha-4\alpha\chi-7)\tau^2+(3+4\alpha\chi^2)\tau-2\alpha\chi^2=0\,,
\end{equation}
and solving for $\alpha$ gives
\begin{equation}
\alpha=
\frac{(4\tau^3-7\tau^2+3\tau)}{\big[(4\chi-2)\tau^2-4\chi^2\tau+2\chi^2\big]}\,.
\end{equation}
\end{subequations}
These loci for $\alpha$ are plotted as a function of $\tau$ for various
values of $\chi$ in Fig.~\ref{Atf}(a).  For given $\alpha$ and $\chi$, the
extrema occur at the $\tau$ values where the line $\alpha=\mathrm{const.}$
intersects the branches of the curve for a given $\chi$.  For example, in
Fig.~\ref{Atf}(a), when $\alpha=0.5$ and $\chi=0.9$, $\mathcal{F}$ has three
extrema at $\tau=0.32$, 0.64, and 0.99 (two maxima and an intermediate
minimum), as indicated by the dots.  On the other hand for $\alpha=1.25$ and
$\chi=0.7$, there is a single extremum (circle).  When $\chi\approx 0.75$,
the regime of bimodality extends over the widest range of $\alpha$. We plot
the phase boundary (Fig.~\ref{Ac-1st-free}) on the $\chi$-$\alpha$ plane by
scanning across $\chi$ and numerically finding the range of $\alpha$ values at
which three extrema occur.

The behavior near the critical point $(\tau,\alpha)=(\frac{3}{4},1)$ is
particularly intriguing (Fig.~\ref{Atf}(b)).  For a given value of $\chi$
that is less than but very close to $\chi={3}/{4}$, there are two distinct
sets of solutions---one set with $\alpha$ slightly less than 1 and another
set with $\alpha$ slightly larger than 1.  There is also a small gap between
these two solution sets, as indicated in Fig.~\ref{Atf}(b) for the cases
$\chi=0.749$ and $\chi=0.7499$.  The consequence of this feature is that the
unimodal to bimodal phase boundary is not single valued near the cusp, as
shown in Fig.~\ref{Ac-1st-free}.  It is quite remarkable that such a simple
question about a Brownian path---namely the time for the path to cross an
intermediate position---leads to such a rich phenomenon.

\section{Intermediate Crossings for First-Passage Paths}
\label{sec:FPP}

We now turn to the properties of intermediate crossings for first-passage
paths, where the first-passage constraint for the full trajectory from
$(0,0)$ to $(X,T)$ significantly affects the global nature of intermediate
trajectory and concomitantly, the intermediate crossings.  We divide our
discussion into: (i) the location of the particle at an arbitrary
intermediate time and (ii) the time when the particle crosses an arbitrary
intermediate position.

\subsection{Intermediate Crossing Position}

At a time $t<T$, we again ask: what is the probability $\mathcal{P}(x,t)$
that the particle is at position $x$, given that the particle first reaches
$X$ at time $T$?  We determine this probability, following the same approach
that led to Eq.~\eqref{CP}, by decomposing the trajectory into the segment up
to time $t$ and the remaining segment from $T-t$ to $T$.  The first segment
is a Brownian path to $(x,t)$, but now subject to the constraint that it can
not reach $X$, because reaching $X$ would mean that the full trajectory is
not a first-passage path to $(X,T)$.  To implement this constraint we impose
an absorbing boundary condition at $X$ by augmenting the initial Gaussian
with an anti-Gaussian image distribution that is centered at
$2X$~\cite{F68,R01}.  We denote the probability distribution in the presence
of this absorbing boundary condition as $P_A(x,t)$.  The second segment from
$T-t$ to time $T$ is simply a first-passage path between these two points.

Assembling these elements, we have
\begin{align}
\label{CPF}
  \mathcal{P}(x,t)&= \frac{ P_A(x,t)\,\,
               F(X\!-\!x,T\!-\!t)}{F(X,T)}\nonumber \\
     &= \frac{\displaystyle{
            \frac{1}{\sqrt{4\pi Dt}}\left[e^{-x^2/4Dt}\!-\!e^{-(2X\!-\!x)^2/4Dt}\right]
            \frac{\left\vert X\!-\!x\right\vert }{\sqrt{4\pi D(T\!-\!t)^3}}\, e^{-(X\!-\!x)^2/4D(T\!-\!t)}}}
             {\displaystyle{  \frac{\left\vert X\right\vert }{\sqrt{4\pi DT^3}}\, e^{-X^2/4DT}}}\,.
\end{align}
By expressing all quantities in terms of the dimensionless variables
introduced previously, the above expression simplifies to
\begin{align}
\label{Pct}
\mathcal{P}(\chi,\tau) &= \sqrt{\frac{\alpha}{\pi}} \frac{\left\vert 1-\chi\right\vert }{\sqrt{\tau(1-\tau)^3}} \,\,
e^{-\alpha\big[(1-\chi)^2-(1-\tau)\big]/(1-\tau)}
\left[e^{-\alpha\chi^2/\tau}-e^{-\alpha(2-\chi)^2/\tau}\right]\nonumber\\
&=  \sqrt{\frac{\alpha}{\pi}} \frac{\left\vert 1-\chi\right\vert }{\sqrt{\tau(1-\tau)^3}} \,\,
e^{-\alpha(\chi-\tau)^2/[\tau(1-\tau)]}\left[1-e^{-4\alpha(1-\chi)/\tau}\right]\,.
\end{align}
Notice that $\tau$ lies in $(0,1)$ by construction, but $\chi$ can range from
$-\infty$ to 1.

\begin{figure}[ht]
\centerline{\includegraphics[width=1.05\textwidth]{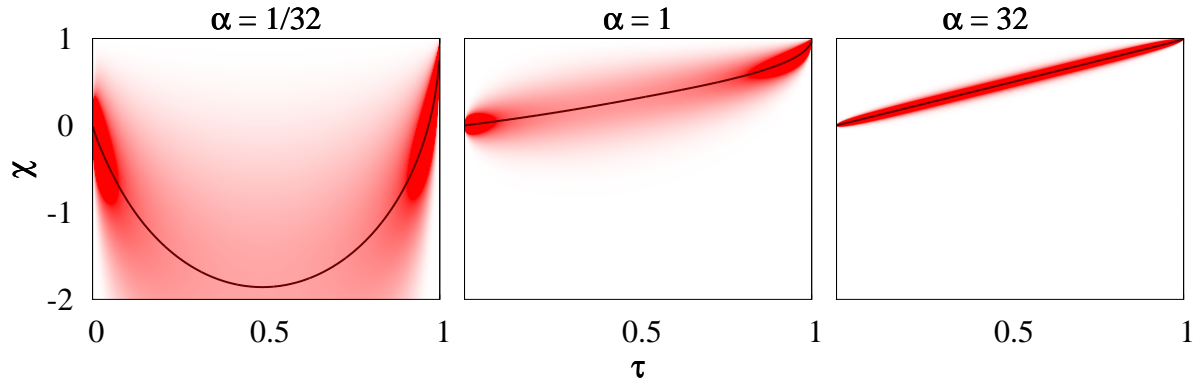}}
\caption{Heatmaps of the conditional occupation probability
  $\mathcal{P}(\chi,\tau)$ for a first-passage path from $(0,0)$ to $(X,T)$
  for illustrative values of $\alpha$ in the $\tau$-$\chi$ plane.  The smooth
  curve shows the time dependence of the most probable location of the
  particle.  For small $\alpha$, the trajectory is initially repelled from
  its final destination.  The distribution is most diffuse near
  $\tau=\frac{1}{2}$ and most localized near $\tau=0$ and $\tau=1$.}
\label{figCP}
\end{figure}

The behavior of $\mathcal{P}(\chi,\tau)$ is quite rich, as illustrated in
Fig.~\ref{figCP}.  For small $\alpha$ (diffusive limit), the particle must
initially move in the negative-$x$ direction so as to ensure that it does not
hit $X$ before time $T$.  A related type of repulsion arises in the positions
of a set of $N$ random walkers that all start a finite distance away from an
absorbing point subject to the constraint that none of them are absorbed up
to a given time~\cite{MR14}.  As $\alpha$ is increased, corresponding to a
more ballistic and hence more deterministic trajectory from $(0,0)$ to
$(X,T)$, the spread in the probability distribution correspondingly
decreases.  

The extent of the repulsion may be quantified by the minimum value of
$x_{\rm mp}(\tau)$, the most probable location of the trajectory as a
function of rescaled time.  For $\alpha\to 0$, the difference between the
initial and final positions becomes immaterial so that we can approximate the
minimum value of $x_{\rm mp}(\tau)$ by $x_{\rm mp}(\tau\!=\!\frac{1}{2})$.
Using this simplification in Eq.~\eqref{Pct}, the probability distribution at
$\tau=\frac{1}{2}$ reduces to
\begin{equation}
\label{Pcts}
\mathcal{P}(y,\tfrac{1}{2}) \propto \sqrt{\alpha}\,e^{-\alpha}\,y\,\,
e^{-4\alpha y^2} \sinh(4\alpha y)\,,
\end{equation}
where $y=\chi-1$.  Finding the maximum of this latter expression is now
elementary and we find that for $\alpha\to 0$, the minimum on the
most probable trajectory is located at
\begin{equation}
y*\simeq -\frac{1}{2\sqrt{\alpha}}~.
\end{equation}
Thus for $\alpha\to 0$, the most probable trajectory is strongly repelled
from the final point for $\tau<1/2$ and subsequently is strongly attracted to
this final point.  We note that this behavior for $y*$ can also be inferred
from a macroscopic fluctuation theory, as given in Ref.~\cite{MR14}.

\subsection{Intermediate Crossing Times}

Parallel to the discussion given for the pure Brownian path, we now compute
the distribution of times when the first-passage trajectory to $(X,T)$
crosses an intermediate level $x \in (0,X)$. We shall consider the case $x<0$
separately in section~\ref{sec:neg} as the calculation details are different.
We also separately consider the cases of the first crossing and the last
crossing because the calculational details and the results for these two
cases are quite different.

\subsubsection{First-crossing probability}

As in the case where the full trajectory is an unconstrained Brownian path
that starts at $(0,0)$ and ends at $(X,T)$, we decompose the trajectory into
an initial segment that starts at $(0,0)$ and crosses $x$ for the first time
at time $t$, and a final segment that starts at $(x,t)$ and crosses $X$ for
the first time at time $T$.  We consider here the case $x>0$; the case $x<0$
will be treated in the next section.  Here the analog of Eq.~\eqref{CF} is
\begin{equation}
\label{CFF}
\mathcal{F}(x,t) = \frac{F(x,t)\,F(X-x,T-t)}{F(X,T)}\,.
\end{equation}
We now substitute the relevant first-passage probabilities from \eqref{PF}
into the above equation to obtain
\begin{align}
\mathcal{F}(x,t) =   \frac{\left\vert  x\right\vert }{\sqrt{4\pi D t^3}}
\,\, e^{-x^2/4Dt}
\frac{\left\vert  X-x\right\vert }{\sqrt{4\pi D(T-t)^3}}
\,\,e^{-(X-x)^2/4D(T-t)} \!\Bigg/ \!
\frac{\left\vert  X\right\vert }{\sqrt{4\pi DT^3}} 
\,\,e^{-X^2/4DT}.
\end{align}
In terms of the dimensionless variables $\alpha$, $\chi$, and $\tau$, the
above expression simplifies, after some straightforward algebra, to
\begin{equation}
\label{firstcrossing}
\mathcal{F}(\chi,\tau) = \sqrt{\frac{\alpha}{\pi}} \; 
\frac{\left\vert  1-\chi\right\vert   \left\vert \chi\right\vert }{\big[(1-\tau)\tau\big]^{3/2}}\; 
e^{-\alpha(\tau-\chi)^2/[\tau(1-\tau)]}\,.
\end{equation}

\begin{figure}[ht]
\centerline{\includegraphics[width=0.9\textwidth]{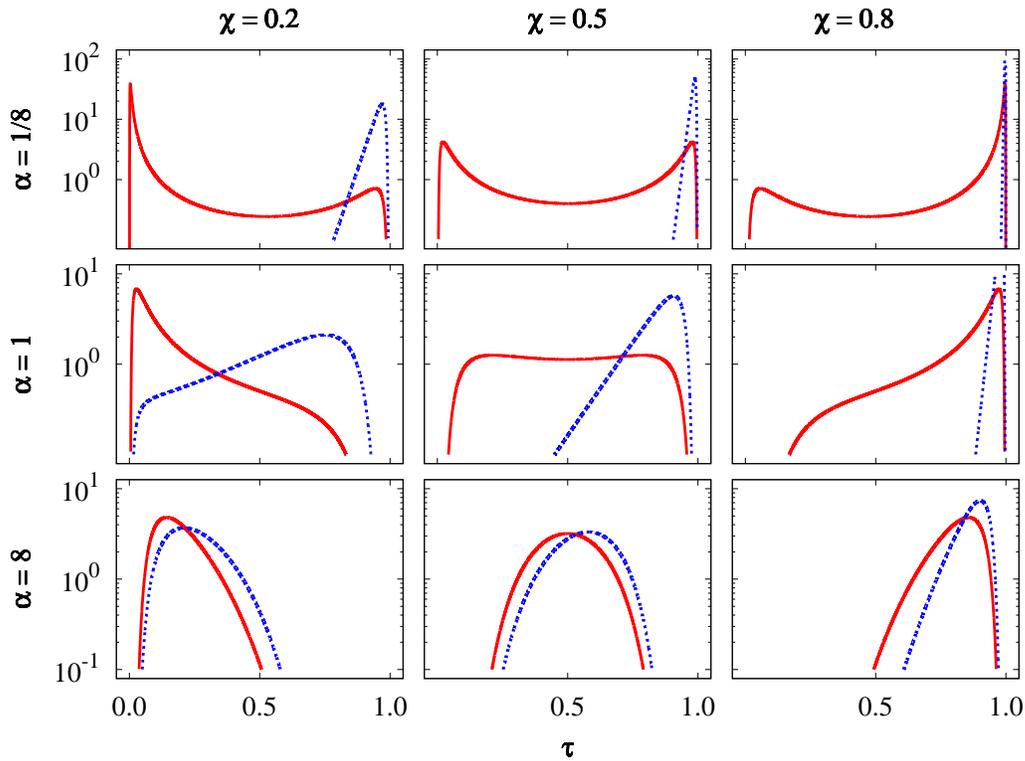}}
\caption{ First-crossing probability $\mathcal{F}(\chi,\tau)$ (red, solid)
  and the last-crossing probability $\mathcal{L}(\chi,\tau)$ (blue, dashed)
  for illustrative values of $\alpha$ and $\chi$ when the initial trajectory
  is a first-passage path from $(0,0)$ to $(X,T)$.  For small $\alpha$,
  $\mathcal{F}(\chi,\tau)$ is bimodal while $\mathcal{L}(\chi,\tau)$ is
  unimodal and sharply peaked as $\tau\to 1$.  For large $\alpha$ (almost
  ballistic trajectory), the first-crossing and last-crossing times nearly
  coincide. }
\label{CF-CL-2}
\end{figure}

Once again, the qualitative behavior of this first-crossing probability
depends in an essential way on the value of $\alpha$ (Fig.~\ref{CF-CL-2}).
For large $\alpha$ (ballistic limit) $\mathcal{F}(\chi,\tau)$ is sharply
peaked about the point $\tau=\chi$.  Thus in this ballistic limit, the most
probable time for a Brownian particle to first reach a distance $x$ is simply
equal to $x$.  Moreover, the distribution in \eqref{firstcrossing} reduces to
a Gaussian form for $\tau\approx\chi$.

Conversely, in the diffusive limit of $\alpha\to 0$, and for $\tau\gg \alpha$
and $1-\tau\gg \alpha$ (i.e., $\tau$ not too close to $0$ or $1$), the
probability density function is controlled by the factor
$\big[\tau(1-\tau)\big]^{-3/2}$, which is similar to, but more singular than
the arcsine law~\cite{MP10}.  The surprising result for this limit is that a
Brownian particle is most likely to cross an arbitrary intermediate level
either at the very beginning or at the very end of its trajectory.

\begin{figure}[ht]
\centerline{\resizebox{0.75\textwidth}{!}{\input{figs/symm.pspdftex}}}
\caption{(a) A first-passage path that consists of a solid and dashed segment
  with a first-crossing at $(\chi,\tau)$.  Interchanging the order of these
  segments gives a first-crossing at $(1\!-\!\chi,1\!-\!\tau)$. }
\label{symm}
\end{figure}
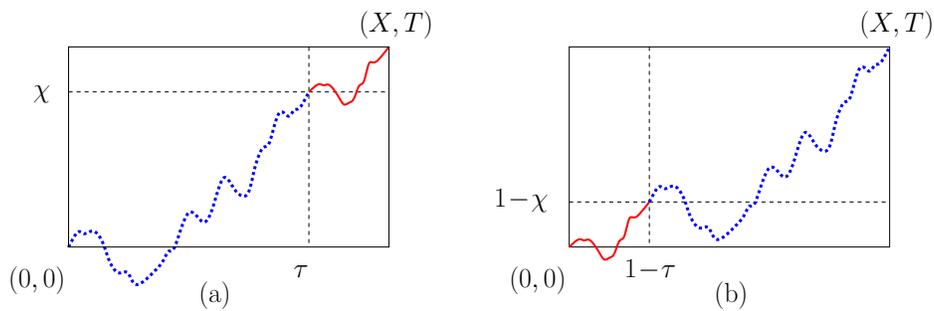

A remarkable aspect of the first-crossing probability \eqref{firstcrossing}
is its invariance under the simultaneous interchanges $\tau\to 1\!-\!\tau$
and $\chi\to 1\!-\!\chi$.  We can give a simple graphical argument to justify
this symmetry (Fig.~\ref{symm}).  In (a), a first-passage path from $(0,0)$
to $(X,T)$ is comprised of a first crossing (dashed) to $(\chi,\tau)$ (in
scaled units) and the remaining segment to $(X,T)$ (solid).  Interchanging
these two segments leads to a first-crossing segment to
$(1\!-\!\chi,1\!-\!\tau)$ and the remaining segment to $(X,T)$.  Since the
segments are independent, the probability for these two first-passage paths
in the figure are identical and thus
$\mathcal{F}(\chi,\tau) = \mathcal{F}(1\!-\!\chi,1\!-\!\tau)$.

\begin{figure}[ht]
\centerline{\resizebox{0.75\textwidth}{!}{\input{figs/bimod.pspdftex}}}
\caption{(a) First-passage paths that first cross an intermediate level near
  $\tau=0$ or $\tau=1$ on an exaggerated scale to emphasize the limit
  $\alpha\ll 1$.  (b) A first-passage path that first crosses the
  intermediate level near $\tau=\frac{1}{2}$.}
\label{bimod}
\end{figure}
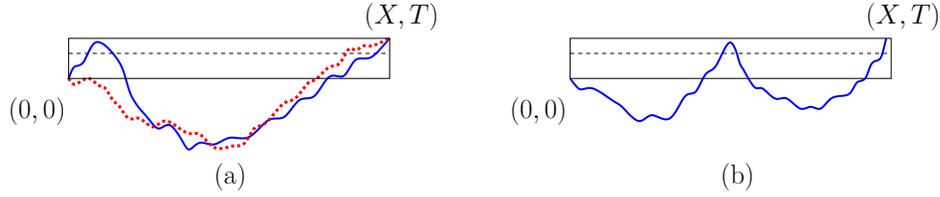

We can apply this same perspective to argue that $\mathcal{F}$ is bimodal for
$\alpha\ll 1$.  Indeed, let us compare a first-passage path to the final
point that has its first crossing close to $\tau=0$ or $\tau=1$, and a
first-passage path that has its first crossing near $\tau=1/2$.  In the
former case (see Fig.~\ref{bimod}(a)), the remainder of the path away from
the first crossing must be repelled from the final point.  The probability
for such a first-passage path from $(0,0)$ to $(X,T)$ scales as $T^{-3/2}$
for $\alpha\ll 1$.  On the other hand, the probability for a path that
first-crosses $\chi$ near $T/2$ is the product of the probabilities of two
first-passage paths of duration $T/2$, namely $\big(T^{-3/2}/2\big)^2$, which
is much less than $T^{-3/2}$.  Thus a first-crossing near $T/2$ is unlikely
for $\alpha\ll 1$.

\begin{figure}[ht]
\subfigure[]{\includegraphics[width=0.48\textwidth]{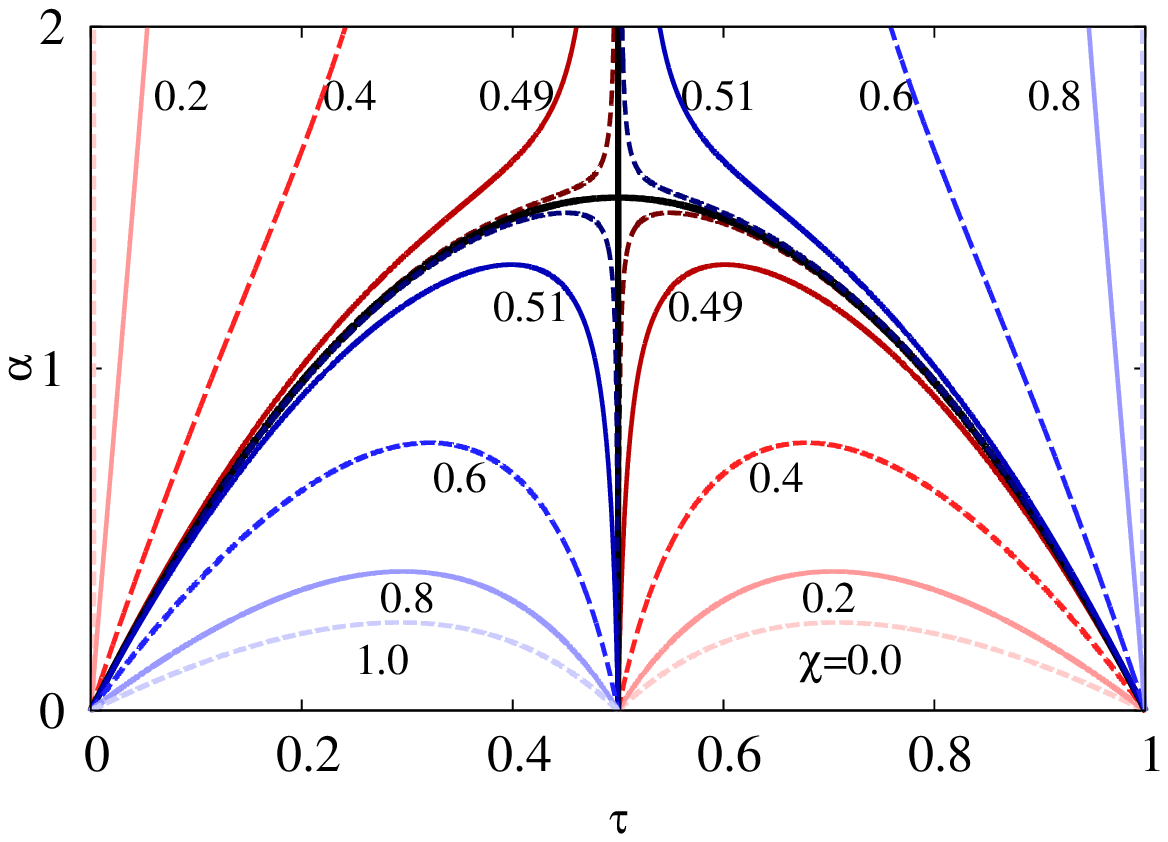}}
\subfigure[]{\includegraphics[width=0.48\textwidth]{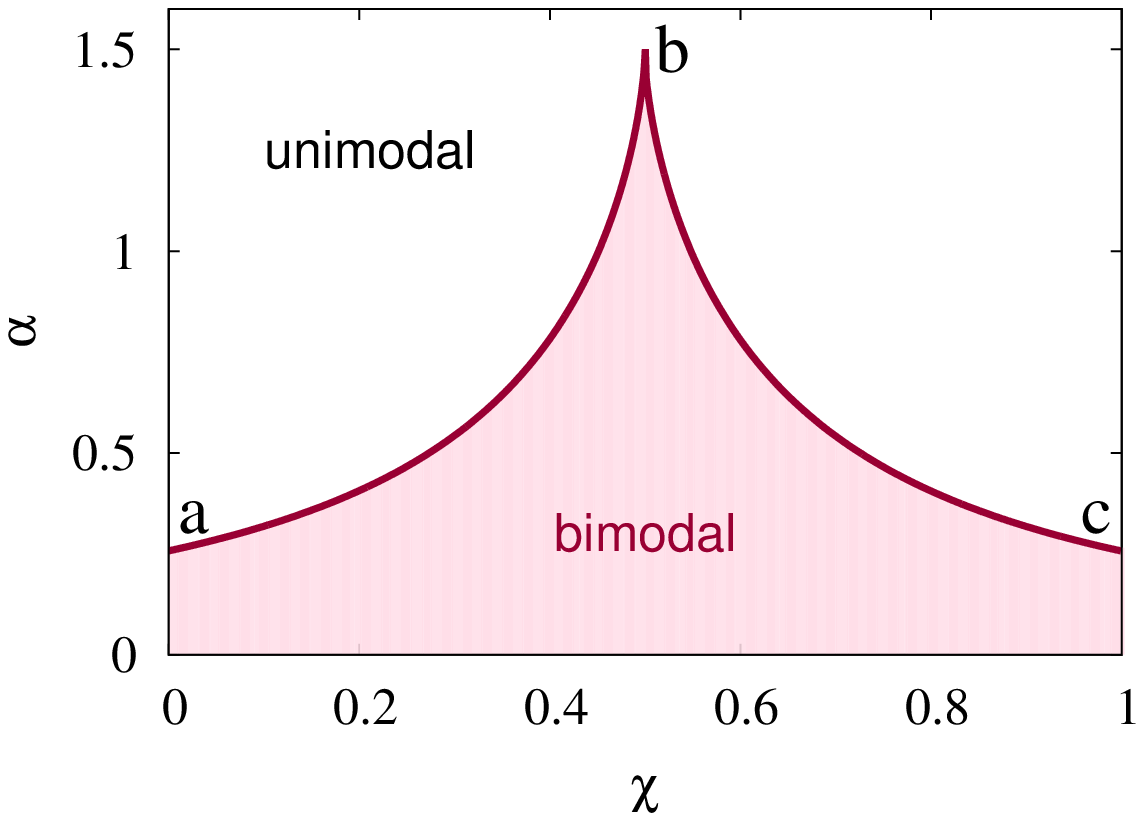}}
\caption{(a) Loci of the extrema of $\mathcal{F}(\chi,\tau)$ in the
  $\alpha$-$\tau$ plane for various values of $\chi$ when the initial
  trajectory is a first-passage path to $(X,T)$.  For a given $\alpha$ and
  $\chi$, the extrema occur at the values of $\tau$ where the line
  $\alpha=\mathrm{const.}$ intersects the curve that corresponds to a given
  $\chi$. (b) Numerically obtained phase diagram from Eq.~\eqref{cubic-fpp},
  showing where in the parameter space the first-crossing probability, for a
  first-passage Brownian path from $(0,0)$ to $(X,T)$, is unimodal and where
  it is bimodal.  The points marked $a,b,c$ are at
  $(0,\frac{9}{2}-3\sqrt{2})$, $(\frac{1}{2},\frac{3}{2})$, and
  $(1,\frac{9}{2}-3\sqrt{2})$, respectively.}
\label{Atc}
\end{figure}

To determine where in the phase space the transition in $\mathcal{F}$ between
unimodality and bimodality occurs, we again determine the extrema of
$\mathcal{F}$.  Following the same procedure that led to Eq.~\eqref{extr}, we
obtain the cubic equation for the location of the extrema
\begin{equation}
\label{cubic-fpp}
6\tau^3 +(2\alpha-4\alpha\chi-9)\tau^2+(3+4\alpha\chi^2)\tau-2\alpha\chi^2=0\,,
\end{equation}
and the resulting solutions for $\alpha$ are plotted as a function of $\tau$
for various values of $\chi$ in Fig.~\ref{Atc}(a).  For given $\alpha$ and
$\chi$, the extrema occur at the $\tau$ values where the line
$\alpha=\mathrm{const.}$ intersects the curve corresponding to a given
$\chi$.  Figure~\ref{Atc}(b) shows the region of the phase diagram where the
first-crossing probability is unimodal and where it is bimodal.  Here the
phase boundary is everywhere single valued, in contrast to the case of
Fig.~\ref{Ac-1st-free} the initial path is an unrestricted Brownian motion.
As a result of the symmetry of $\mathcal{F}$, the phase diagram is symmetric
about the point $\chi=\frac{1}{2}$. Also, unlike in the case of unconstrained
Brownian path, the phase boundary is single-valued.

\subsubsection{Last-crossing probability}

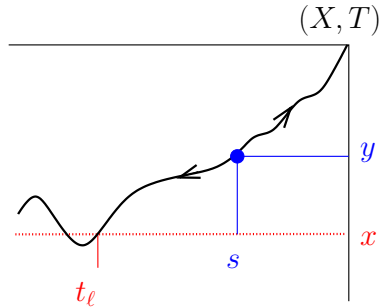
\begin{figure}[ht]
\centerline{\resizebox{0.3\textwidth}{!}{\input{figs/2-segs.pspdftex}}}
\caption{Zoom of the first-passage path in Fig.~\ref{model}, with the
  decomposition of the segment from $t_\ell$ to $T$ into a time-reversed
  first-passage path from $(y,s)$ to $(x,t_\ell)$ and a first-passage path
  from $(y,s)$ to $(X,T)$. }
\label{2-segs}
\end{figure}

We now determine when the particle crosses a specified intermediate level for
the \emph{last} time.  In principle, we may perform this calculation by
decomposing the trajectory into an initial segment from $(0,0)$ to the last
crossing of $x$ at time $t$, and a remaining first-passage segment from
$(x,t)$ to $(X,T)$.  The subtle feature here is that this second segment must
also obey the constraint that this segment always remains greater than $x$,
so that no additional crossings of $x$ occur after time $t$.  To satisfy the
latter condition, we must also impose an absorbing boundary condition at $x$.
Thus the particle begins the second segment at the absorbing boundary, so
that the first-passage probability to $(X,T)$ would equal zero.  To sidestep
this pathology, we could start the particle at $x+dx$ and take the limit $dx$
at the end of the calculation.  This limiting process is a delicate,
however, and we therefore give an alternative approach that avoids any
limiting processes.  The price, however, is the necessity to break the
trajectory into three segments (Fig.~\ref{2-segs}).

In this alternative approach, the first segment is a Brownian path that
starts at $(0,0)$ and reaches the last crossing of $x$ at time $t_\ell$.  An
absorbing boundary at $X$ must be imposed to ensure that this segment never
reaches $X$ before $T$.  We now break the remaining segment from $(x,t_\ell)$
to $(X,T)$, into two sub-segments with respect to an arbitrary point $(y,s)$,
with $x<Y<X$ and $t_\ell<s<T$.  The left sub-segment is a backward-propagating
first-passage path from $(y,s)$ to $(x,t_\ell)$ and the right sub-segment is a
forward-propagating first-passage path from $(y,s)$ to $(X,T)$.  The choice
of intermediate point $(y,s)$ is arbitrary, so that the final result must be
independent of $s$ after integrating over all $y$.

Thus we have
\begin{equation}
\label{CL2}
\mathcal{L}(x,t)=\frac{P_A(x,t) \,\int_x^X F_A(y-x,s-t) F_A(X-y,T-s) dy}{F(X,T)}\,.
\end{equation}
Here $F_A(z,t)$ denotes the first-passage probability in the presence of the
absorbing boundary condition at $X$.  Making use of standard results for the
first-passage probability in the interval~\cite{R01}, the integral in
\eqref{CL2} is
\begin{align}
\label{CL2-int}
\int_x^X & F_A(y-x,s-t) F_A(X-y,T-s) dy \nonumber \\
&\hskip -0.5cm = \int_x^X \!\frac{4\pi^2 D^2}{(X\!-\!x)^4} 
\!\sum_{m,n=1}^\infty\!  (-1)^m n m \sin\!\left(\frac{n\pi y}{X\!-\!x}\right)
\sin\!\left(\frac{m\pi y}{X\!-\!x}\right) e^{-\pi^2 D[n^2 (s-t)+m^2 (T-s)]/(X-x)^2}dy \nonumber \\
&\hskip -0.5cm = \frac{2\pi^2 D^2}{(X-x)^3} \sum_{n=1}^\infty (-1)^n n^2 e^{-n^2\pi^2 D (T-t)/(X-x)^2}\,.
\end{align}
To obtain the third line, the orthogonality relation
$\int_0^1 2 \sin(n\pi x)\sin(m\pi x) dx = \delta_{m n}$ has been used.

\begin{figure}
\centerline{\subfigure[]{\includegraphics[width=0.5\textwidth]{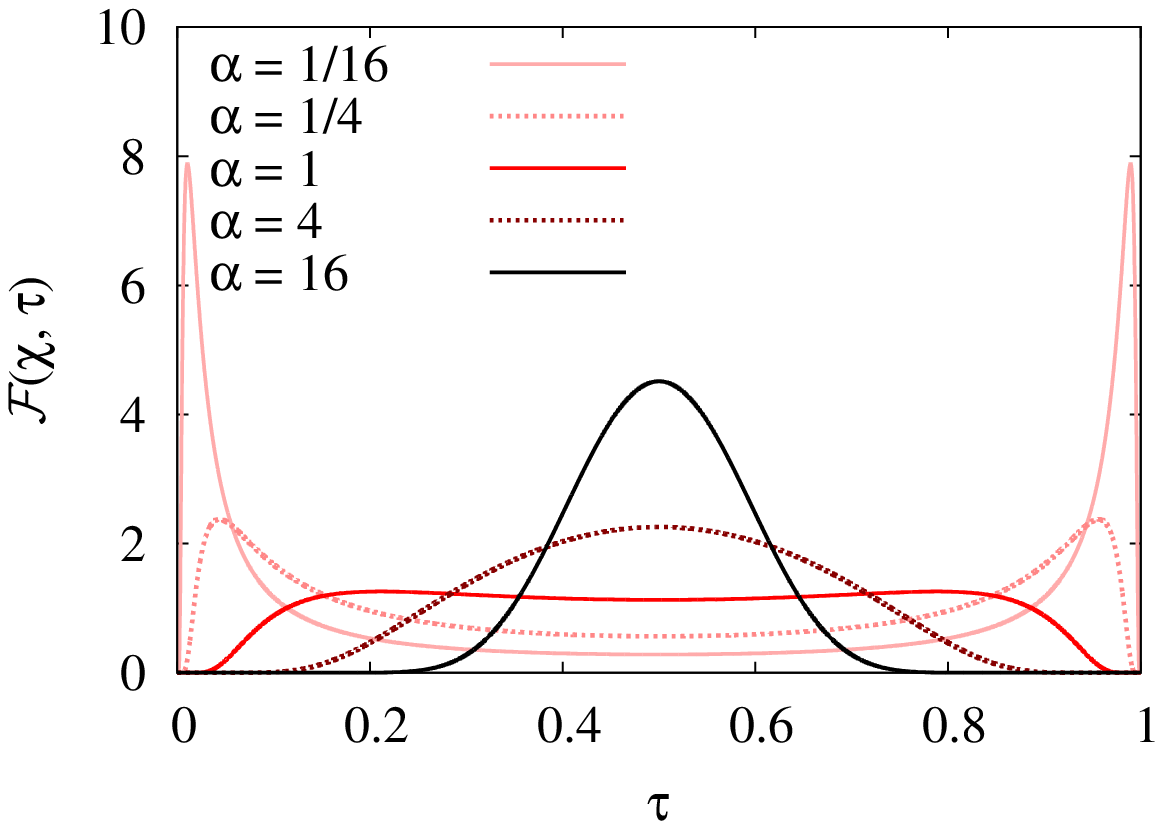}}
\subfigure[]{\includegraphics[width=0.5\textwidth]{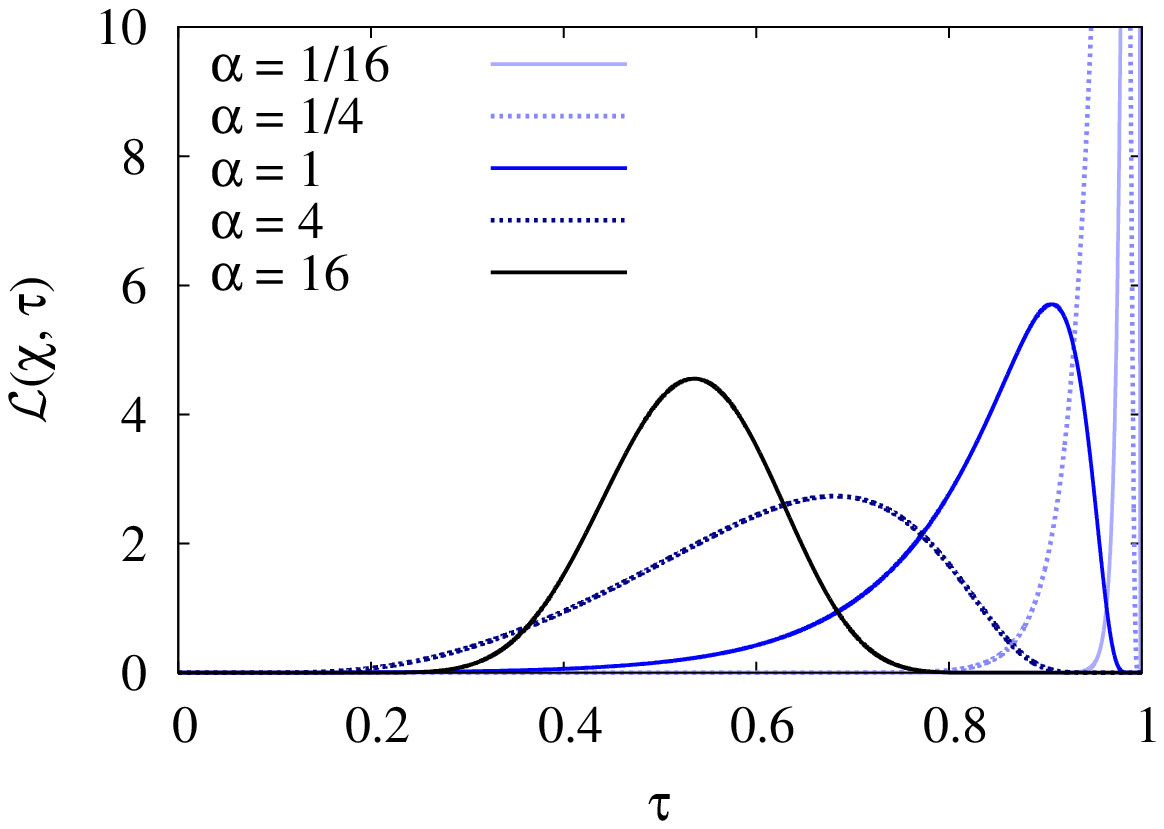}}}
\caption{ (a) Comparison of the first-crossing probability
  $\mathcal{F}(\chi,\tau)$ and (b) the last-crossing probability
  $\mathcal{L}(\chi,\tau)$ for representative values of $\alpha$ for the
  specific case of $\chi= 1/2$.  As $\alpha$ increases
  $\mathcal{F}(\chi,\tau)$ remains symmetric, but changes from bimodal to
  unimodal, while $\mathcal{L}(\chi,\tau)$ merely moves systematically toward
  $\tau=1$.}
\label{pdfs-first-last}
\end{figure}

Substituting the above expression, together with the previously derived forms
for $P_A(x,t)$ and $F(X,T)$ in Eq.~\eqref{CL2}, and using the dimensionless
variables $\chi$, $\tau$ and $\alpha$, we finally obtain
\begin{equation}
\label{L-final}
\mathcal{L}(x,t) = \Big(\frac{\pi^2 e^\alpha}{8\alpha^2\sqrt{\tau}(1\!-\!\chi)^3}\Big)
\left[e^{-\alpha\chi^2/\tau}\!-\!e^{-\alpha(2-\chi)^2/\tau}\right]
\left[\sum(-1)^n n^2 e^{-n^2\pi^2 (1-\tau)/[4\alpha(1-\chi)^2]}\right] \,.
\end{equation}
When $\alpha \ll 1$, for $1-\tau\gg\alpha$ ($\tau$ not too close to $1$), the
first term in the sum dominates and the last-crossing probability reduces to
\begin{equation}
\mathcal{L}(\chi,\tau) \sim \frac{e^{-A(1-\tau)}}{\tau^{1/2}}\,,
\end{equation}
with $A = \pi^2/[4\alpha(1-\chi)^2]$.  In this limit, the last-crossing
probability increases exponentially with $\tau$ as $\tau\rightarrow 1$
(Fig.~\ref{pdfs-first-last}).  Conversely, when $\alpha \gg 1$, the
last-crossing probability is sharply peaked at $\tau = \chi$.  This feature
is a result of the trajectory becoming nearly ballistic.  When
$\alpha \gg 1$, the first and last crossing times coincide because there is
little stochasticity in the trajectory.

\section{First and Last Crossings to $x<0$}
\label{sec:neg}

Thus far, we have tacitly assumed that the intermediate position is between 0
and $X$.  With this constraint, every path from $(0,0)$ to $(X,T)$ must
necessarily pass through any intermediate position.  We now investigate the
first- and last-crossing probabilities when the intermediate position $x<0$.
Here the probability that a first-passage path from $(0,0)$ to $(X,T)$
actually reaches $x$ is less than one.  The decomposition of the first
passage path into a segment from the start to the crossing point and from the
crossing point to the final point still applies, but there are additional
subtleties associated with the intermediate point being negative.

Again, we separately consider the first- and last-crossing probabilities.  It
is convenient to define $z=\vert x\vert $, which is manifestly positive.  The
distribution of first-crossing times to $z$ is formally given by
\begin{align}
\label{CFz}
\mathcal{F}(z,t) &= \frac{F_A(z,t)\,F(X+z,T-t)}{F(X,T)}\,.
\end{align}
The constituent first-passage probability $F_A(z,t)$ is subject to the
constraint that $X$ cannot be reached before time $t$; this requirement
ensures that the full path is actually a first-passage trajectory to $(X,T)$.
The second segment from $(z,t)$ to $(X,T)$ is a first-passage path between
these two points without any additional constraints.

To obtain $F_A(z,t)$, we again solve the diffusion equation in the interval
between $z=-x$ and $X$, with absorbing boundary conditions at both ends, and
then compute the flux to the boundary point $z$ for a particle that starts at
the origin.  By standard methods~\cite{R01}, this computation gives
\begin{align}
F_A(z,t) & = \frac{2\pi D}{(X+z)^2} \sum_{n=1}^\infty n\,\sin{\left(\frac{n\pi z}{X+z}\right)}\,\,
 e^{-n^2\pi^2 D t/(X+z)^2}\,.
\end{align}
Thus the first-crossing probability is
\begin{equation} 
\mathcal{F}(z,t)  =
\frac{\displaystyle{ \frac{2\pi D}{(X\!+\!z)^2} \sum_{n=1}^\infty n 
\sin{\left(\frac{n\pi z}{X+z}\right)}\,
 e^{-n^2\pi^2 D t/(X+z)^2}
 \frac{\left\vert  X \!+\! z\right\vert }{\sqrt{4\pi D (T\!-\! t)^3}}
 e^{-(X+z)^2/4D(T-t)}}}
{\displaystyle \frac{\left\vert  X\right\vert }{\sqrt{4 \pi  D T^3}} e^{-X^2/4DT}}
\label{below-origin-first-1}
\end{equation}
In dimensionless variables, the above expression simplifies to
\begin{equation}
\mathcal{F}(\chi,\tau) = 
\frac{\pi e^{\alpha[1-\tau-(1+\chi)^2]/(1-\tau)}}{2\alpha
  (1+\chi)(1-\tau)^{3/2}}
\sum_{n=1}^\infty n \sin{\left(\frac{n\pi\chi}{1+\chi}\right)} 
e^{-n^2\pi^2\tau/[4\alpha(1+\chi)^2]}\,.
\end{equation}

\begin{figure}
  \centerline{
\subfigure[]{\includegraphics[width=0.5\textwidth]{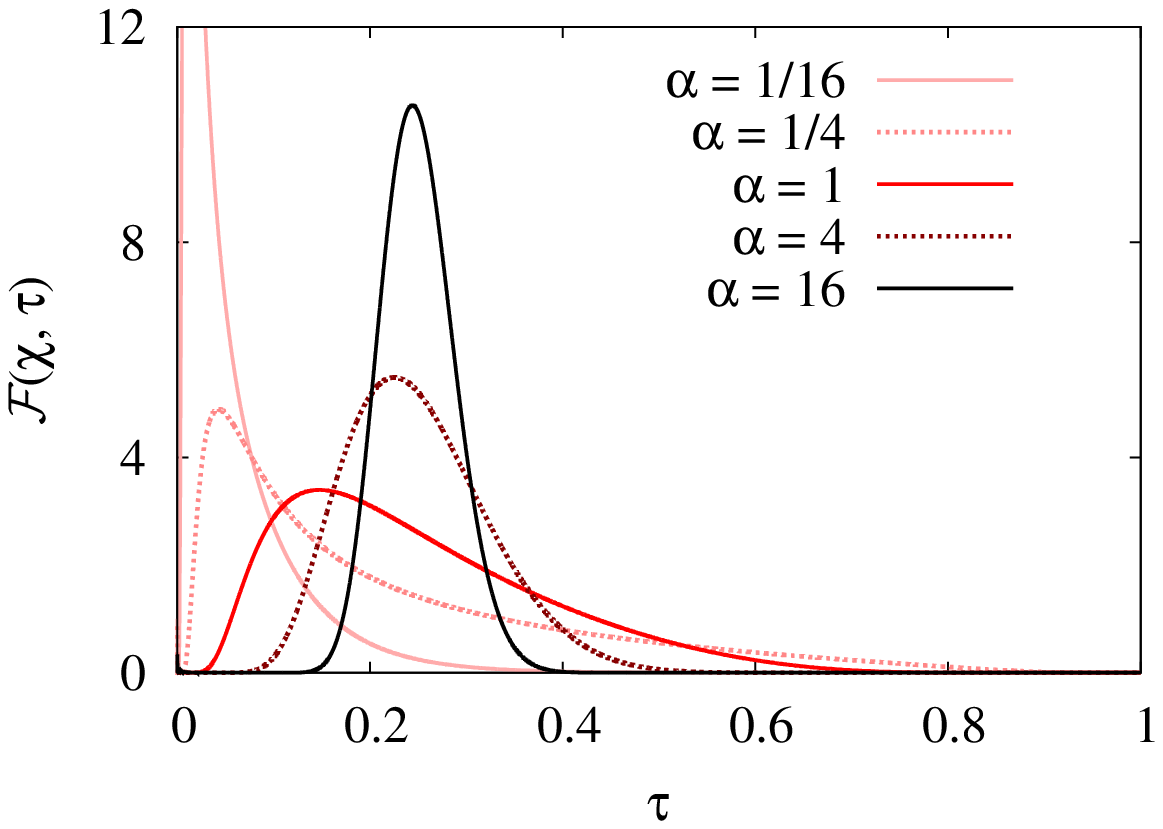}}
\subfigure[]{\includegraphics[width=0.5\textwidth]{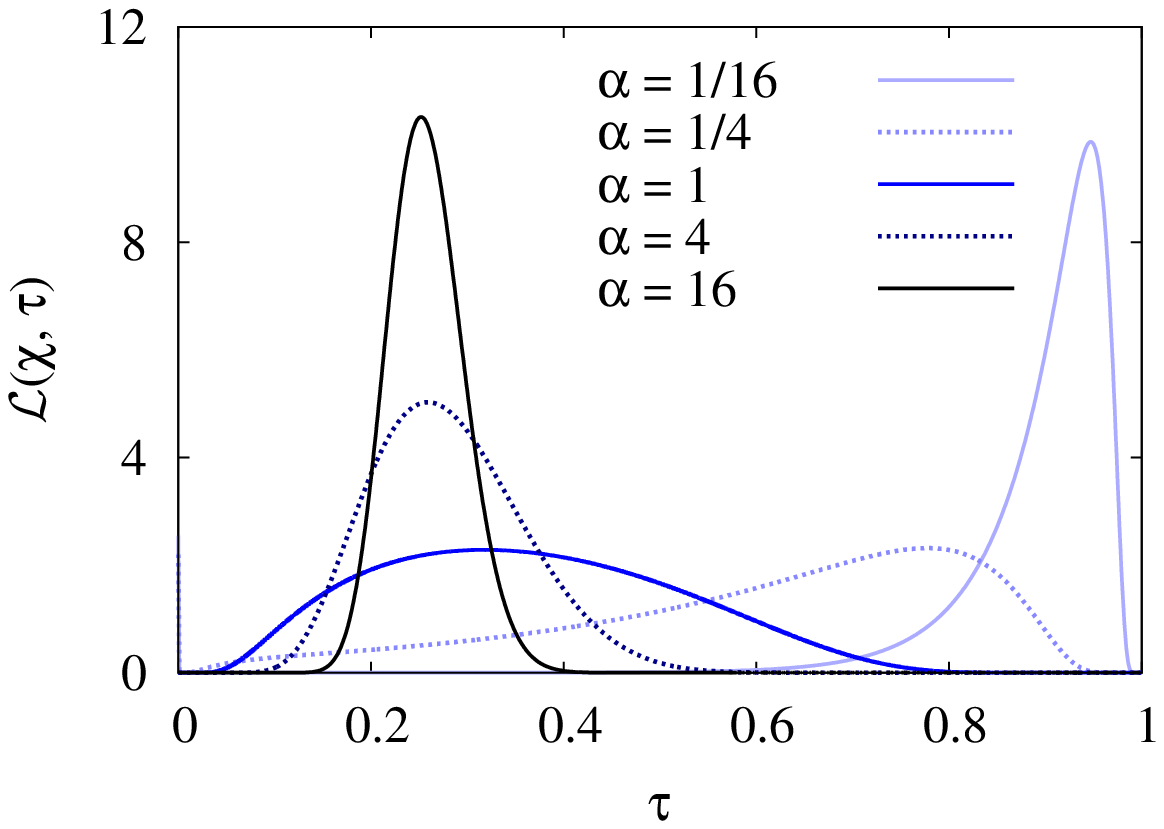}}}
\caption{ (a) The first-crossing probability $\mathcal{F}(\chi,\tau)$ and (b)
  the last-crossing probability $\mathcal{L}(\chi,\tau)$ for
  $\chi=-1/2$ for a range of $\alpha$ values.  The distributions have been
  rescaled so that they fit on the same range.}
\label{FL-neg}
\end{figure}

For $\alpha \rightarrow 0$, for $\tau\gg\alpha$, the first term in the sum is dominant so that 
\begin{equation}
  \mathcal{F}(\chi,\tau) \sim \frac{e^{-A\tau}}{(1-\tau)^{3/2}}\qquad \alpha\to
  0\,,
\end{equation}
with $A=\pi^2/\big[4\alpha (1+\chi)^2\big]$.  Conversely, for
$\alpha \rightarrow \infty$, Fig.~\ref{FL-neg} shows that
$\mathcal{F}(\chi,\tau)$ is peaked at $\tau=\frac{\chi}{1+\chi}$.  The
particle thus moves ballistically in the negative-$x$ direction to reach $-x$
in a fraction $\frac{\chi}{1+\chi}$ of the total time and then moves
ballistically in the positive-$x$ direction to reach $X$ at time $T$.

The distribution of last-crossing times has the identical form to that given
in Eq.~\eqref{L-final} for the case where $x>0$, except for the substitution
$x\to -x$.  For $\alpha\ll 1$, the last crossing necessarily occurs very
close to the final time $\tau=1$.  For $\alpha\gg 1$, the overall trajectory
is a straight line in the negative-$x$ direction to the intermediate point
and then another straight line in the positive-$x$ direction to the final
point.  In this limit, the first- and last-crossing probabilities become
progressively more similar, as expected intuitively.

\section{Summary and Outlook}
\label{sec:summ}

We investigated the properties of intermediate crossings of Brownian and
first-passage paths in one dimension that start at $(0,0)$ and end at a final
point $(X,T)$.  By using simple probabilistic arguments, we analytically
determined the occupation probability at a general intermediate time, as well
as the first-crossing and last-crossing times at a general intermediate
location.  These intermediate properties exhibit a number of surprising and
anomalous features that depend, in an essential way, on $\alpha=X^2/4DT$.  In
the ballistic limit of $\alpha\gg 1$, a trajectory typically moves
systematically from the starting point to the final point.  In this case, one
can infer intermediate-crossing properties by linearly interpolating the
initial trajectory.

The more interesting situation of the diffusive limit, where $\alpha\ll 1$, a
first-passage Brownian path to $(X,T)$ must initially move away from this
final point so as to not reach $X$ before time $T$.  The maximum excursion
away from the starting point scales as $1/\sqrt{\alpha}$.  The behavior of
the first-crossing probability, the distribution of times when a path first
crosses a specified intermediate point $x$ is quite rich.  When the
ostensibly simpler example where initial path is a Brownian trajectory to
$(X,T)$ with no additional constraint, the first-crossing probability can be
either unimodal or bimodal, with bimodality favored for small $\alpha$ and
$\chi=x/X\approx 0.75$.  Moreover, the phase boundary between unimodality and
bimodality is, quite surprisingly, not single valued.  When the initial path
is a first-passage path to $(X,T)$, the first-crossing probability again has
a unimodal to bimodal transition and the corresponding phase diagram is
symmetric in the $\alpha$-$\chi$ plane.

Another unanticipated feature of the first- and last-crossing probabilities
is that the results are invariant if there is an overall global drift in the
Brownian trajectory.  When one generalizes the expressions for the first- and
last-crossing probabilities given in Eqs.~\eqref{CFL}, \eqref{CFF}, and
\eqref{CL2} to incorporate a constant drift, all factors that involve the
drift velocity cancel out.  It would be worthwhile to understand the full
ramifications of this simple observation.

Finally, it is worth mentioning that the decomposition of the first-passage
path in \eqref{CFF} provides the starting point for efficient simulations of
intermediate crossing phenomena.  The naive way to numerically determine
intermediate crossing properties is by direct simulation of a random walk in
which the times at which various intermediate positions are reached are
recorded for a walk that reaches a given final point.  From this data, one
can reconstruct the first-crossing (as well as the last-crossing)
probabilities.  As a much more efficient alternative, one can first select a
set of intermediate positions and then directly move a particle only between
these intermediate positions.  Correspondingly, the time between each of
these macro-steps is incremented from the appropriate distribution of first
passage times.  This procedure can be made still more efficient by only
allowing the walk to move to intermediate positions that are progressively
closer to the final position.  Thus arbitrarily long random walks can be
simulated with a finite number of steps.

We thank Satya Majumdar for helpful suggestions and literature advice.
Financial support for this research was provided in part by grant No.\
2012145 from the United States-Israel Binational Science Foundation (BSF),
Grant No.\ DMR-1205797 from the National Science Foundation, and by a grant
from the John Templeton Foundation.

\end{document}

%% file: figs/model.pspdftex
\begin{picture}(0,0)%
\includegraphics{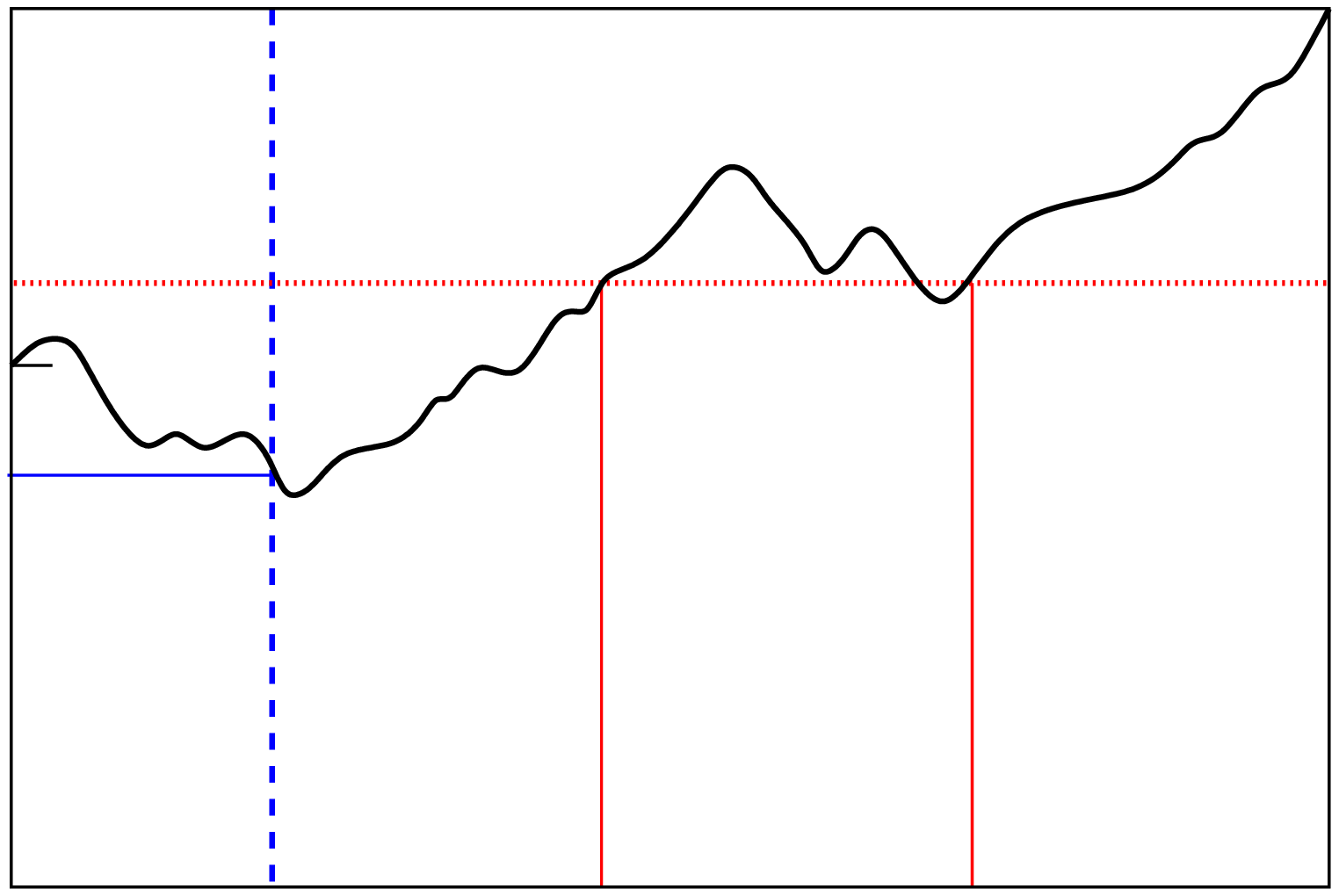}%
\end{picture}%
\setlength{\unitlength}{3947sp}%
\begingroup\makeatletter\ifx\SetFigFont\undefined%
\gdef\SetFigFont#1#2#3#4#5{%
  \reset@font\fontsize{#1}{#2pt}%
  \fontfamily{#3}\fontseries{#4}\fontshape{#5}%
  \selectfont}%
\fi\endgroup%
\begin{picture}(8288,6265)(1346,-8267)
\put(7426,-8086){\makebox(0,0)[lb]{\smash{{\SetFigFont{34}{40.8}{\familydefault}{\mddefault}{\updefault}{\color[rgb]{1,0,0}$t_\ell$}%
}}}}
\put(9151,-2461){\makebox(0,0)[lb]{\smash{{\SetFigFont{34}{40.8}{\familydefault}{\mddefault}{\updefault}{\color[rgb]{0,0,0}$(X,T)$}%
}}}}
\put(1361,-4861){\makebox(0,0)[lb]{\smash{{\SetFigFont{34}{40.8}{\familydefault}{\mddefault}{\updefault}{\color[rgb]{0,0,0}$(0,0)$}%
}}}}
\put(1801,-5421){\makebox(0,0)[lb]{\smash{{\SetFigFont{34}{40.8}{\familydefault}{\mddefault}{\updefault}{\color[rgb]{0,0,1}$x$}%
}}}}
\put(5401,-8086){\makebox(0,0)[lb]{\smash{{\SetFigFont{34}{40.8}{\familydefault}{\mddefault}{\updefault}{\color[rgb]{1,0,0}$t_f$}%
}}}}
\end{picture}%

%% file: figs/symm.pspdftex
\begin{picture}(0,0)%
\includegraphics{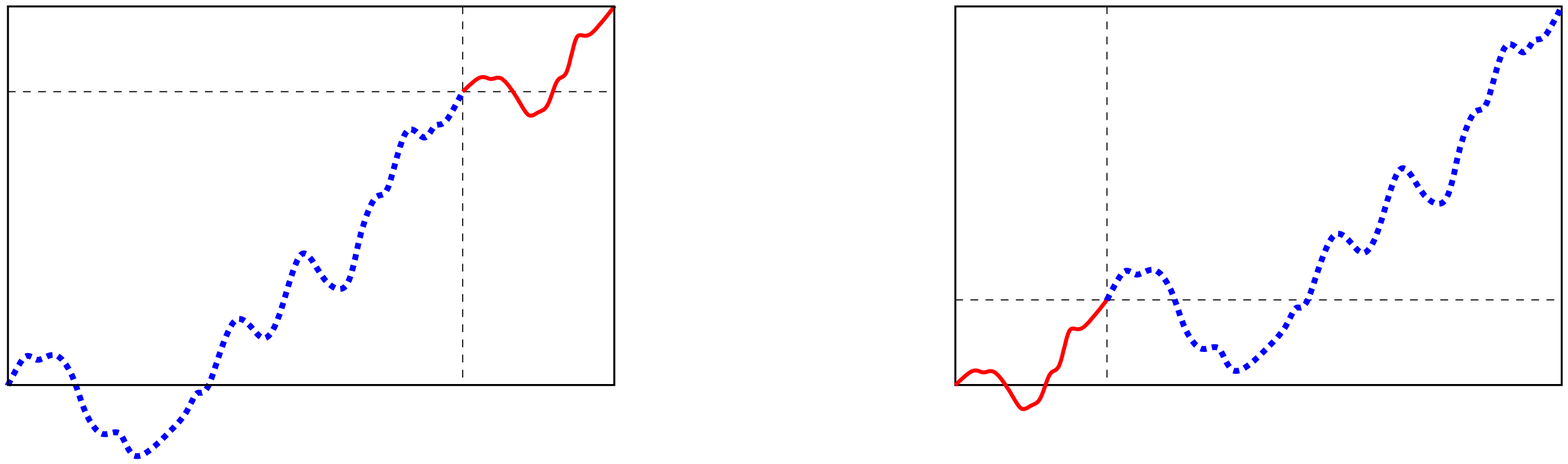}%
\end{picture}%
\setlength{\unitlength}{3947sp}%
\begingroup\makeatletter\ifx\SetFigFont\undefined%
\gdef\SetFigFont#1#2#3#4#5{%
  \reset@font\fontsize{#1}{#2pt}%
  \fontfamily{#3}\fontseries{#4}\fontshape{#5}%
  \selectfont}%
\fi\endgroup%
\begin{picture}(13259,4579)(286,-5540)
\put(9526,-5011){\makebox(0,0)[lb]{\smash{{\SetFigFont{29}{34.8}{\familydefault}{\mddefault}{\updefault}{\color[rgb]{0,0,0}$1\!-\!\tau$}%
}}}}
\put(5551,-1336){\makebox(0,0)[lb]{\smash{{\SetFigFont{29}{34.8}{\familydefault}{\mddefault}{\updefault}{\color[rgb]{0,0,0}$(X,T)$}%
}}}}
\put(301,-5161){\makebox(0,0)[lb]{\smash{{\SetFigFont{29}{34.8}{\familydefault}{\mddefault}{\updefault}{\color[rgb]{0,0,0}$(0,0)$}%
}}}}
\put(3151,-5386){\makebox(0,0)[lb]{\smash{{\SetFigFont{29}{34.8}{\familydefault}{\mddefault}{\updefault}{\color[rgb]{0,0,0}(a)}%
}}}}
\put(13051,-1336){\makebox(0,0)[lb]{\smash{{\SetFigFont{29}{34.8}{\familydefault}{\mddefault}{\updefault}{\color[rgb]{0,0,0}$(X,T)$}%
}}}}
\put(7801,-5161){\makebox(0,0)[lb]{\smash{{\SetFigFont{29}{34.8}{\familydefault}{\mddefault}{\updefault}{\color[rgb]{0,0,0}$(0,0)$}%
}}}}
\put(10876,-5386){\makebox(0,0)[lb]{\smash{{\SetFigFont{29}{34.8}{\familydefault}{\mddefault}{\updefault}{\color[rgb]{0,0,0}(b)}%
}}}}
\put(676,-2311){\makebox(0,0)[lb]{\smash{{\SetFigFont{29}{34.8}{\familydefault}{\mddefault}{\updefault}{\color[rgb]{0,0,0}$\chi$}%
}}}}
\put(4576,-5011){\makebox(0,0)[lb]{\smash{{\SetFigFont{29}{34.8}{\familydefault}{\mddefault}{\updefault}{\color[rgb]{0,0,0}$\tau$}%
}}}}
\put(7576,-3961){\makebox(0,0)[lb]{\smash{{\SetFigFont{29}{34.8}{\familydefault}{\mddefault}{\updefault}{\color[rgb]{0,0,0}$1\!-\!\chi$}%
}}}}
\end{picture}%

%% file: figs/bimod.pspdftex
\begin{picture}(0,0)%
\includegraphics{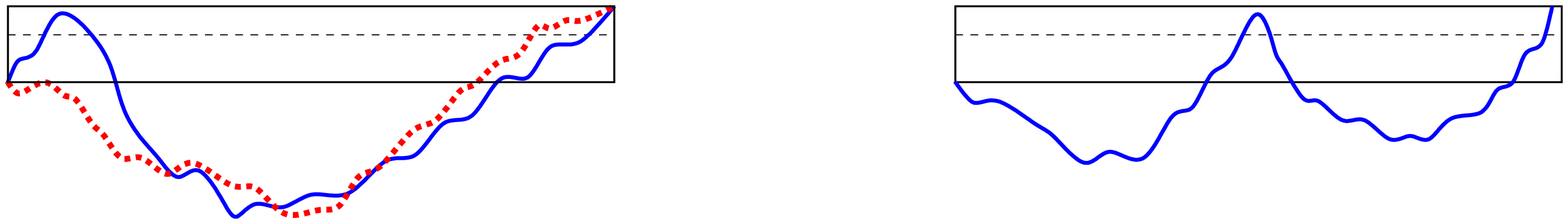}%
\end{picture}%
\setlength{\unitlength}{3947sp}%
\begingroup\makeatletter\ifx\SetFigFont\undefined%
\gdef\SetFigFont#1#2#3#4#5{%
  \reset@font\fontsize{#1}{#2pt}%
  \fontfamily{#3}\fontseries{#4}\fontshape{#5}%
  \selectfont}%
\fi\endgroup%
\begin{picture}(13237,2929)(286,-6290)
\put(10951,-6136){\makebox(0,0)[lb]{\smash{{\SetFigFont{29}{34.8}{\familydefault}{\mddefault}{\updefault}{\color[rgb]{0,0,0}(b)}%
}}}}
\put(301,-5161){\makebox(0,0)[lb]{\smash{{\SetFigFont{29}{34.8}{\familydefault}{\mddefault}{\updefault}{\color[rgb]{0,0,0}$(0,0)$}%
}}}}
\put(7801,-5161){\makebox(0,0)[lb]{\smash{{\SetFigFont{29}{34.8}{\familydefault}{\mddefault}{\updefault}{\color[rgb]{0,0,0}$(0,0)$}%
}}}}
\put(13126,-3736){\makebox(0,0)[lb]{\smash{{\SetFigFont{29}{34.8}{\familydefault}{\mddefault}{\updefault}{\color[rgb]{0,0,0}$(X,T)$}%
}}}}
\put(5626,-3736){\makebox(0,0)[lb]{\smash{{\SetFigFont{29}{34.8}{\familydefault}{\mddefault}{\updefault}{\color[rgb]{0,0,0}$(X,T)$}%
}}}}
\put(3376,-6136){\makebox(0,0)[lb]{\smash{{\SetFigFont{29}{34.8}{\familydefault}{\mddefault}{\updefault}{\color[rgb]{0,0,0}(a)}%
}}}}
\end{picture}%

%% file: figs/2-segs.pspdftex
\begin{picture}(0,0)%
\includegraphics{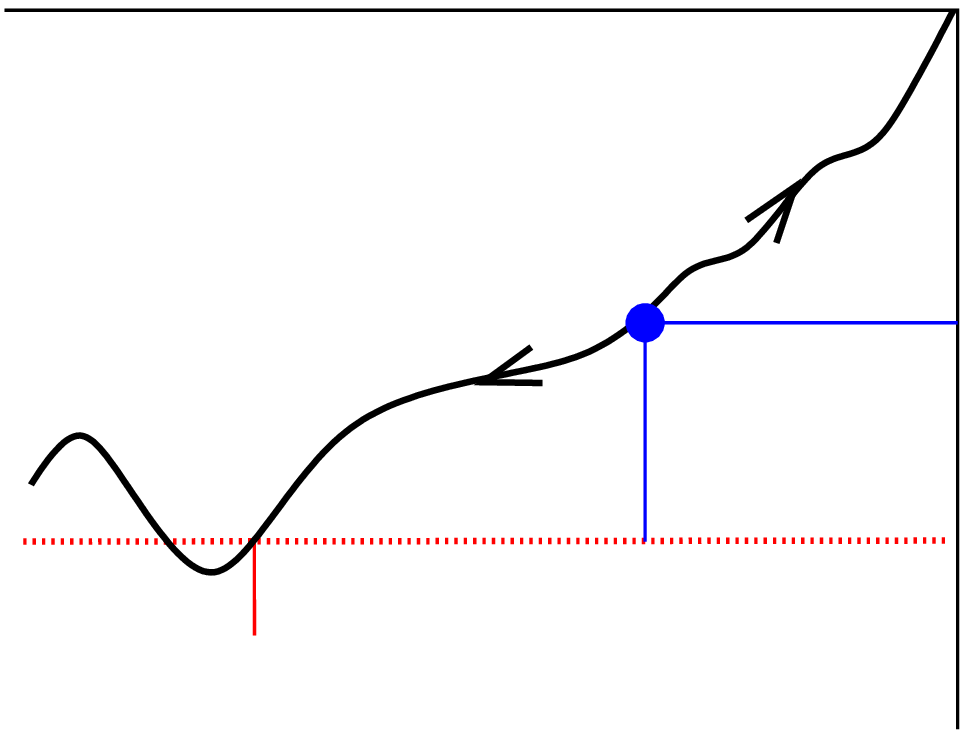}%
\end{picture}%
\setlength{\unitlength}{3947sp}%
\begingroup\makeatletter\ifx\SetFigFont\undefined%
\gdef\SetFigFont#1#2#3#4#5{%
  \reset@font\fontsize{#1}{#2pt}%
  \fontfamily{#3}\fontseries{#4}\fontshape{#5}%
  \selectfont}%
\fi\endgroup%
\begin{picture}(4762,4216)(6354,-4115)
\put(7276,-3961){\makebox(0,0)[lb]{\smash{{\SetFigFont{25}{30.0}{\familydefault}{\mddefault}{\updefault}{\color[rgb]{1,0,0}$t_\ell$}%
}}}}
\put(11101,-2011){\makebox(0,0)[lb]{\smash{{\SetFigFont{25}{30.0}{\familydefault}{\mddefault}{\updefault}{\color[rgb]{0,0,1}$y$}%
}}}}
\put(9301,-3511){\makebox(0,0)[lb]{\smash{{\SetFigFont{25}{30.0}{\familydefault}{\mddefault}{\updefault}{\color[rgb]{0,0,1}$s$}%
}}}}
\put(11101,-3211){\makebox(0,0)[lb]{\smash{{\SetFigFont{25}{30.0}{\familydefault}{\mddefault}{\updefault}{\color[rgb]{1,0,0}$x$}%
}}}}
\put(10276,-286){\makebox(0,0)[lb]{\smash{{\SetFigFont{25}{30.0}{\familydefault}{\mddefault}{\updefault}{\color[rgb]{0,0,0}$(X,T)$}%
}}}}
\end{picture}%